\documentclass[12pt]{iopart}

\usepackage{graphics}
\usepackage{iopams}

\newcommand{\half}{\mbox{\scriptsize$1\over2$}}
\newcommand{\quart}{\mbox{\scriptsize$1\over4$}}

\newcommand{\imi}{\mathrm{i}}
\newcommand{\dex}{\mathbin{\mathrm{d}}}

\newcommand{\ph}{\varphi}
\newcommand{\xx}{\tilde x}
\newcommand{\yy}{\tilde y}
\newcommand{\vv}{\check v}

\newcommand{\myselector}[2]{{%
\ifx\letterversion\undefined%
{#2}%
\else%
{#1}%
\fi
}}

\newcommand{\Eq}{{%
\ifx\letterversion\undefined
Eq.\
\fi
}}

\newcommand{\Eqs}{{%
\ifx\letterversion\undefined
Eqs.\
\fi
}}

\newcommand{\maddr}{
Mendeleevo, Moscow Region,\\ 
141570, Russia
}

\newcommand{\abstr}{
The first order nonlinear ODE 
$\dot \ph(t) + \sin\ph(t)=q(t),q(t)=B+A\cos\omega t,$ 
where $ A,B,\omega$ are real constants,
is considered, the transformation converting it to
a second order linear homogeneous
ODE with polynoimial coefficients is found.
The latter is identified as a particular case of the double
confluent Heun equation.
The series of algebraic constraints on the constant parameters is found 
whose fulfillment leads to the existance of solutions representable
through polynomials in explicit form.
These polynomials are found to constitute
the orthogonal normalizable system.
}

\begin{document}

\ifx\letterversion\undefined
\title[Modelling of Josephson junction]%
{The modelling of a Josephson junction
and Heun polynomials}%
\else
\letter{
The modelling of a Josephson junction
and Heun polynomials
}
\fi

\author{S.I. Tertychniy}

\ifx\letterversion\undefined
\address{VNIIFTRI, Russia}
\else
\address{\maddr}%
\ead{XXXX@mail.ru}
\fi

\ifx\letterversion\undefined
\begin{abstract}
\abstr
\end{abstract}
\else
\relax
\fi

\ifx\letterversion\undefined
\relax
\else
\fi


\ifx\letterversion\undefined
\relax
\else
\submitto{\JPA}
\fi

\bigskip

Nowadays, electronic devices based on 
the Josephson effect in superconductors and, in particular,
Josephson junctions (JJ)  \cite{A1}
play the important role 
in the measurement technique, 
serving, in particular,
the core element of the modern voltage standards \cite{A2}.
The application needs lead to the growing importance
of the theoretical and mathematical modelling of  JJ properties.
One of the theoretical tools commonly
used for this purpose
is
the RSJ (Resistively Shunted Junction) 
model  \cite{A3,A3b}
which applies, in the case of {\em overdamped\/} JJ \cite{A3c}, 
the ODE
 \begin{equation}
 \dot \ph(t) + \sin\ph(t)=q(t).
                                  \label{eq:1}
\end{equation}
Here the (real valued) function $\ph(t)$ called {\it the phase\/}
and describing  JJ state
is unknown while $q(t)$
representing the external impact to  JJ
(the appropriately normalized {\it bias current} supplied by an external source)
is assumed to be given.
The dot denotes the derivative with respect to the variable $t$
 (the appropriately normalized current time).

The goal of the present 
\myselector%
{Letter}%
{notes}
is 
the \myselector{presentation}{discussion}
of some results concerning
the equation 
(\ref{eq:1}) and its solutions in
the particular case
of harmonic $q(t)$ most important for applications.
Thus we assume,
without loss of generality,
 \begin{equation}\label{eq:1a} 
 q(t)=B+A\cos\omega t,
\end{equation}
where  $A,B,\omega$ are some constants subject to the
condition  $A\not=0\not=\omega$ in order
to eliminate trivial situations.

As it was first noted by V.M.\ Buchstaber (see \cite{A4}), 
the first order nonlinear 
ODE (\ref{eq:1}) is equivalent, for arbitrary $q(t)$,
to the system of two linear  ODEs
\begin{equation}\label{eq:2}
\eqalign{
2\dot x(t)=\hphantom{-[} x(t)+q(t) y(t)\hphantom{]},
                                                \nonumber\\
2\dot y(t)=- [q(t) x(t)+ y(t)],}
\end{equation}
where $x(t),y(t)$ are 
the new unknowns.
They are connected with $\varphi(t)$ by the equation 
\begin{equation}\label{eq:3}
  \exp(\imi\ph(t))={x(t)-\imi y(t)\over x(t)+\imi y(t)}.
\end{equation}
\myselector{\null}%
{Thus $\varphi$ is twice the phase of the complex quantity $x-\imi y$.}

In the case of $q(t)$ defined by \Eq (\ref{eq:1a}),
the 
replacing of the independent real variable $t$ 
with the complex valued variable
\begin{equation}\label{eq:4}
  z=\exp(\imi\omega t),
\end{equation}
translates \Eqs (\ref{eq:2})  to
the equations
\begin{eqnarray}\label{eq:5}
\eqalign{
4\imi\omega z \;\xx(z)'
=2 \xx(z)
+\left[2B+A\left(z+z^{-1}\right)\right]  \yy(z),
                                                \cr
4\imi\omega z\; \yy(z)'
=
-\left[2B+A\left(z+z^{-1}\right)\right]  \xx(z)- 2\yy(z).
}%
\end{eqnarray}
Here the separate notations for the unknowns $x,y$ considered 
as functions of the complex variable $z$,
 $\xx(z)=x(t),\yy(z)=y(t)$,
are employed, the prime denotes the derivative with
respect to $z$.
\myselector{\null}{
Multiplying them by $z$, the equations with polynomial coefficients result.}

One may extend the original meaning of $\xx,\yy$,  treating them as analytic 
functions of the complex variable $z$
which
satisfy \Eq (\ref{eq:5})%
\myselector{.}{
 everywhere in the complex plane except singular points.
 Then, obviously, a non-zero
 real or imaginary parts of any pair 
 of such functions which prove smooth on some segment of
 the unit circle in $\mathbb C$ 
 yield a {\it real\/ } solution of \Eqs (\ref{eq:2})
 and, consequently, a (real) solution of \Eqs 
 (\ref{eq:1}),(\ref{eq:1a}).
}

Now let us consider the following transformation 
replacing the unknowns $\xx,\yy$ with the functions $v=v(z),\vv=\vv(z)$
in accordance with equations
\begin{equation}
  \label{eq:6}
\eqalign{
  v= \hphantom{(2\omega z)^{-1} }
\llap{\mbox{$\imi \:$}}
 z^{-{B\over2\omega}}
\exp
\left({A\over4\omega}\left(-z+z^{-1}\right)
\right)
(\xx-\imi\yy),
\\
\vv=(2\omega z)^{-1}
 z^{-{B\over2\omega}}
\exp
\left({A\over4\omega}\left(-z+z^{-1}\right)
\right)
(\xx+\imi\yy).}
\end{equation}
It is easy to show that 
the fulfilment of 
\Eqs
(\ref{eq:5}) is equivalent to the condition of the vanishing of
the following two expressions
\begin{eqnarray}
  \label{eq:7}
&&v'
-\vv
\mbox{\ and\ }
z^2\vv' 
+\left({A\over2\omega}(z^2+1)+\left({B\over\omega}+1\right)z\right)\tilde
v
+{1\over4\omega^2}v
\end{eqnarray}
that, in turn, is equivalent to the fulfilment of the
second order linear ODE with polynomial coefficients
\begin{equation}
  \label{eq:8}
\left[
 z^2{\dex^2\over \dex^2 z} 
+\left({A\over2\omega}(z^2+1)+\left({B\over\omega}+1\right)z\right){\dex\over\dex z} 
+{1\over4\omega^2}
\right]
v =0.
\end{equation}
\myselector{\null}{
The latter has the only two singular points, $z=0$ and $z=\infty$.}

The M\"obius transformations 
\begin{equation}
  \zeta={z+\alpha\over z-\alpha},
\end{equation}
where $\zeta$ is the new independent complex variable and $\alpha$
is an arbitrary non-zero complex number,
leads to representations of (\ref{eq:8}) 
more clearly revealing the symmetry in the roles of 
its singular points.
In particular, in the case $\alpha=\imi$, one gets the equation
\begin{equation}
   \label{eq:8b}
 \left[
  (1-\zeta^2)^2{\dex^2\over \dex \zeta^2} 
 +2\left(
\left( {B\over\omega}-\zeta\right)
(1-\zeta^2)-2\imi{A\over\omega}\zeta
 \right)
{\dex\over\dex \zeta} 
 +{1\over\omega^2}
 \right]
 v =0 
 \end{equation}
which proves to be is a particular instance of the
{\em double confluent Heun equation\/}  (DCHE)  as it is given in \cite{A5},
Eq.\ (4.5.11).  It arises 
for the following set of parameters employed in \cite{A5}:
$a=0,c=-(B\omega^{-1}+1),t=\imi
A(2\omega)^{-1},\tilde\lambda=(2\imi\omega A)^{-1}$.
It is also worth noting that
the canonical DCHE representation (Eq.\ (4.5.1) in \cite{A5})
\myselector{\null}{%
$$
z^2{\dex^2y(z)\over\dex z^2}
+(-z^2+c z+t){\dex y(z)\over\dex z}+(-a z+\lambda)y(z)=0
$$
}%
results from (\ref{eq:8}) after the argument rescaling $z\rightarrow
2\imi\omega A^{-1} z$ 
and corresponds to the parameters
$a=0,c=B \omega^{-1}+1,
t=-(\half A\omega^{-1})^2,
\lambda=\quart o^{-2}$.

At the same time
the most elegant form 
of the  M\"obius-transformed \Eq (\ref{eq:8})
results with $\alpha=1$ in which case one gets
 \begin{equation}
   \label{eq:8a}
 \left[
  (1-\zeta^2){\dex\over \dex \zeta}  (1-\zeta^2){\dex\over \dex \zeta} 
 +2\left(
 {B\over\omega}(1-\zeta^2)-{A\over\omega}(1+\zeta^2)
 \right){\dex\over\dex \zeta} 
 +{1\over\omega^2}
 \right]
 v =0 
 \end{equation}
(see \cite{A6}).
Here the singular points $\zeta=\pm1$ 
are just the images
of the only singular points $z=0,\infty$ of \Eq (\ref{eq:8}).
Hence there are no more singular points
for \Eq   (\ref{eq:8a}) and, in particular,
the point $\zeta=\infty$ is regular.
It is also easy 
to show that the only effect caused by the
transition to the reciprocal variable,
$\zeta\rightarrow 1/\zeta$, 
is the inversing of the sign of the parameter $A$ in  (\ref{eq:8a}).
Another $\zeta$ transformation which obviously
preserves the form of  \Eq   (\ref{eq:8a})
is the reflection $\zeta\rightarrow-\zeta$. It
leads to the combined inversion of parameter signs
$A\rightarrow-A,B\rightarrow-B$. 
\myselector{\null}{
Thus, the signs of the both
 parameters $A$ and $B$ are irrelevant
 in the sense they
 can be (independently) reversed by means of transformations of the
 independent variable $\zeta$.}

Yet another useful representation of  \Eq
(\ref{eq:8}) results from  the 
the replacing  of  the unknown $v$  by the new unknown
$P\equiv P(z)$ by means of the substitution
\begin{equation}
  \label{eq:10}
  v=\exp\left({-{A\over2\omega}z}\right)P.
\end{equation}
It is convenient to represent the resulting equation
as follows 
\begin{equation}
  \label{eq:11}
  z(z P' -nP)'
-\mu z (z P'-n P)
+(\mu-z)P'
+\lambda P=0,
\end{equation}
where 
the following new {\it constant parameters\/} $n,\mu,\lambda$ 
replacing the equivalent triplet $A,B,\omega$
in accordance with definitions
\begin{eqnarray}
  \label{eq:11a}
n&=&-\left({B\over\omega}+1\right),\;
\mu={A\over2\omega},\;
\lambda={1-A^2\over4\omega^2} 
={1\over(2\omega)^2}-\mu^2
\end{eqnarray}
are utilized.

The representation (\ref{eq:11})
enables one to conjecture 
that if the parameter $n$ assumes a
non-negative integer value 
then 
this equation may admit 
{\em polynomial\/}
solutions.
Indeed, for $n=0$, a constant is its solution for arbitrary $\omega$,
provided the constraint $\lambda=0$ is additionally obeyed.
Further, for a positive integer  $n$,
let us apply the ansatz
\begin{eqnarray}
  \label{eq:12}
P= P_n=\sum^n_{k=0} a_k z^{k},
\end{eqnarray}
where the constant coefficients 
$a_k$ (dependent also on $n$) have to be determined.
Substituting $P_n$ into (\ref{eq:11}), one gets the following 
system of $n+1$ linear homogeneous equations
\begin{eqnarray}
\fl
  \label{eq:13}
  0&=&\lambda a_0+\mu a_1,
  \label{eq:13a}\\
\fl
0&=&
\mu(n-k+1) a_{k-1}+(\lambda-k(n-k+1))a_k+\mu(k+1)a_{k+1}\;
\;\mbox{for}\;
k=1,\dots,n-1,
  \label{eq:13b}\\
\fl
0&=&\mu a_{n-1}+(\lambda-n) a_n,
\label{eq:13c}
\end{eqnarray}
for $n=1$ the subsystem (\ref{eq:13b}) being void.
It admits a nontrivial solution if and only if
the determinant $\Delta_n(\lambda,\mu)$
\ifx\letterversion\undefined
of the following $(n+1)\times(n+1)$ dimensional 3-diagonal
matrix
\begin{eqnarray}\fl
  \label{eq:14}
\mbox{\normalsize $
\mathbf{\Phi}=
$}
\\
\fl
&\fl\null\hspace{10ex}
\mbox{\scriptsize$
\left(
    \begin{array}{lllllllll}
\lambda\! &\mu\cdot1 \! &0\! &0\! &\dots\! &0\! &0\! &0\! &0
\\[0.9ex]
\mu\cdot n \! &\lambda-1\cdot n \! & \mu\cdot 2
\! &0 \! &\dots  \! &0 \! &0 \! &0 \! &0 
\\[0.9ex]
 0 \! &\mu\cdot (n-1) \! & \lambda-2\cdot(n-1)  \! & \mu\cdot 3 \! &\dots \! &0 \! &0 \! &0 \! &0 
\\[0.9ex]
 0 \! &0\! &\mu\cdot (n-2) \! & \lambda-3\cdot(n-2)  \! &\dots \! &0 \! &0 \! &0 \! &0 
\\[0.9ex]
\dots \! &\dots \! &\dots \! &\dots \! &\dots \! &\dots \! &\dots \! &\dots \! &\dots
\\[0.9ex]
 0 \! &0 \! &0 \! &0 \! &\dots 
 \! & \lambda-(n-3)\cdot 4 \! & \mu\cdot(n-2) \! & 0\! & 0
\\[0.9ex]
 0 \! &0 \! &0 \! &0 \! &\dots \! &\mu\cdot 3
 \! & \lambda-(n-2)\cdot 3 \! & \mu\cdot(n-1) \! & 0
\\[0.9ex]
 0 \! &0 \! &0 \! &0 \! &\dots \! &0 \! &\mu\cdot 2
 \! & \lambda-(n-1)\cdot 2 \! & \mu\cdot n
\\[0.9ex]
0 \! &0 \! &0 \! &0 \! &\dots \! &0 \! &0 \! &\mu\cdot 1 \! &\lambda-n\cdot 1 
    \end{array}
\right)
$}
&\nonumber
\end{eqnarray}
\else
of the $(n+1)\times(n+1)$ dimensional 3-diagonal
matrix $\mathbf{\Phi}$
with the elements defined as follows
\begin{eqnarray}
   \label{eq:14}
\mathbf{\Phi}_{i j}&=&
(\lambda-j(n+1-j))\delta_{j\;l}
+(n-j)\mu \delta_{j\;l-1}
+j\mu\delta_{j\;l+1},
\end{eqnarray}
where $\delta_{a b}$  denotes Kronecker symbol,
\fi
vanishes. 
\begin{equation}
  \label{eq:14a}
  \Delta_n(\lambda,\mu)\equiv\det\mathbf{\Phi}=0
\end{equation}
is the algebraic equation, of degree $n+1$ in $\lambda$,
constraining the parameters $\mu,\lambda$.
Its fulfilment is the necessary and sufficient condition
for the existing of polynomial solutions of \Eq
(\ref{eq:11}).

If (\ref{eq:14a}) is fulfilled,
the polynomial coefficients $a_k$ can be
calculated as follows.
Having introduced
their appropriately scaled ratios $R_k$ by means of the definition
\begin{equation}
  \label{eq:15a}
R_k={\mu\over k}{a_{k-1}\over a_{k}}
\Rightarrow
a_k=a_n
\mu^{k-n}
{\Gamma(n+1)\over\Gamma(k+1)}\prod_{j=k+1}^n R_j
\;\mbox{for\ }k=1,2,\dots,n-1,
\end{equation}
\Eqs (\ref{eq:13b})
are equivalent to the recurrence relations
\begin{eqnarray}
  \label{eq:16}
R_{k}=1+{\lambda\over k(k-n-1)}+
{\mu^2\over k(k-n-1)R_{k+1}},\; k=1,2,\dots
\end{eqnarray}
specifying, generally speaking, a continued fraction.
However, for positive integer $n$,
it is truncated at the $(n-1)$'th step in view of \Eq
(\ref{eq:13c}) which is equivalent to the equation
\begin{equation}
  \label{eq:16a}
R_n=1-\lambda/n.
\end{equation}
Then, making use of \Eqs (\ref{eq:16}),(\ref{eq:16a}),
one may determine, step by step, all $R_k$. In turn, they determine
coefficients $a_k$
up to an arbitrary common factor.

Further,
let us note that \Eq (\ref{eq:15a}) 
can be represented in the following matrix form
\begin{eqnarray}
  \label{eq:17}
  \left[
    \begin{array}{l}
R_k\\1
    \end{array}
\right]
&=&
(Z_k R_{k+1})^{-1}\mathbf{M}_k
 \left[
    \begin{array}{l}
R_{k+1}\\1
    \end{array}
\right],
\\
\llap{\mbox{where}}\;Z_k&=&k(k-n-1)
,
\nonumber\\
\mathbf{M}_k&=&  \left(
    \begin{array}{ll}
Z_k+\lambda & \mu^2
\\
Z_k & 0
   \end{array}
\right).
 \label{eq:17a}
\end{eqnarray}
\Eq (\ref{eq:17})  can also be interpreted as
a linear map on the projective vector space of 2-element columns
defined up to a nonzero factor.
Iterating it
and making use of (\ref{eq:16a}), one gets
\begin{eqnarray}
  \label{eq:18}
    \left[
    \begin{array}{l}
R_k\\1
    \end{array}
\right]
&=&
\left[\prod^{n-1}_{j=k}Z_{j}
\prod^{n}_{j=k+1}R_{j}\right]^{-1}\cdot
%
\prod^{n-1}_{{\longrightarrow \atop j=k}}
\mathbf{M}_j 
\times
\left[
    \begin{array}{l}
1-{\lambda\over n}\\1
    \end{array}
\right].
\end{eqnarray}
Here the symbol 
$\prod \atop \rightarrow$
denotes the product of matrices, where the factors 
corresponding to
 {\em larger\/}
indices $j$ are situated {\em at right} with respect
to the lower index ones.

\Eqs (\ref{eq:16})-(\ref{eq:18}) determine
the set of coefficients $a_k$ 
and, then, the polynomial $P_n(z)$ 
for {\em arbitrary\/} $\lambda,\mu$ irrespectively of
the fulfillment of \Eq (\ref{eq:14a}).
However, if the latter is not satisfied, such a polynomial 
cannot be a solution of \Eq (\ref{eq:11}). Then
some of
\Eqs (\ref{eq:13a})-(\ref{eq:13c}) have to be not satisfied as well.
Since (\ref{eq:13b}),(\ref{eq:13c}) are automatically obeyed by
the very meaning of \Eqs (\ref{eq:16})-(\ref{eq:18}),
it is the fulfillment of \Eq (\ref{eq:11})
which is
equivalent to the equation 
\begin{eqnarray}
  \label{eq:13aa}
    R_1&=&-{\mu^2\over\lambda},
\end{eqnarray}
which 
is, in turn,  connected with fulfilment of \Eq  (\ref{eq:13a}).
Accordingly,
if $R_1$ is considered as the result of calculation with the
help of 
the formula (\ref{eq:18}) for $k=1$,
the equation
\begin{eqnarray}
 \label{eq:13ab}
\left[
    \begin{array}{l}
1\\
{\mu^2
\over 
\lambda
}
    \end{array}
 \right]^{\mbox{\scriptsize\sf T}}
 \times
 \prod^{n-1}_{{\longrightarrow \atop j=1}}
 \mathbf{M}_j 
 \times
 \left[
     \begin{array}{l}
 1-{\lambda\over n}\\1
     \end{array}
 \right]
 &=&0
\end{eqnarray}
arises as the necessary condition of the fulfilment of \Eqs
 (\ref{eq:13a})-(\ref{eq:13c}).
In view of the close resemblance of
algebraic structures of \Eqs (\ref{eq:14a}) and  (\ref{eq:13ab}), 
it is
 then natural to suppose that 
the left-hand-side 
expression of \Eq (\ref{eq:13ab})
is intimately
connected with $\Delta_n(\lambda,\mu)$.
 It is, indeed, the case, and 
the following representation of  $\Delta_n=\det\mathbf{\Phi}$ through the
product of a finite set of $2\times2$
matrices $\mathbf{M}_j$ (\ref{eq:17a}) takes place:
\begin{eqnarray}
  \label{eq:19}
  \Delta_n(\mu,\lambda)&=&
-\left[\lambda \atop \mu^2 \right]^{\mbox{\scriptsize\sf T}}
\times
\prod^{n-1}_{\longrightarrow\atop j=1}
\mathbf{M}_j
\times
 \left[
    \begin{array}{l}
n- \lambda \\n
    \end{array}
\right].
\end{eqnarray}
Similarly, there is the following explicit
representation of the polynomial coefficients $a_k$ 
through the analogous matrix products:
\begin{eqnarray}
  \label{eq:20}
  a_k
&=&
a_n{(-\mu)^{k-n} \over  k(n+1-k)!}
\left[0 \atop 1 \right]^{\mbox{\scriptsize\sf T}}
\times
\prod^{n-1}_{\longrightarrow\atop j=k}
\mathbf{M}_j
\times
 \left[
    \begin{array}{l}
n-{\lambda}\\n
    \end{array}
\right], \;k=1,\dots,n-1.
\end{eqnarray}
Although, formally, it does not cover the case $k=0$, 
a minor modification allows to compute $a_0$ as well
(one has to replace {\it in the multipliers\/}
the integer parameter $k$ 
with $k+\varepsilon$, where $\varepsilon$ is a small real number,
to carry out computation with $k=0$,
and to pass to the limit $\varepsilon\rightarrow0$
in the result).

The polynomials $P_n$
reveal a remarkable symmetry concerning the
coefficients in front of the ``small''
and ``large'' $z$  powers. It can be 
discovered considering the following polynomial constructed from $P_n$:
\begin{equation}
  \label{eq:21}
  \tilde P_n(z)=z^n
[P'_n(z^{-1})-\mu P_n(z^{-1})].
\end{equation}
A straightforward computation
shows that it 
obeys \Eq (\ref{eq:11}) if and only if
$P_n(z)$ does. 
However, the polynomial solution 
of \Eq (\ref{eq:11}) is unique up to a normalization.
Indeed, the second  solution linearly independent 
with $P_n(z)$ admits the following representation in quadratures:
\begin{equation}
Q_n=P_n\int 
z^n
\exp\left(\mu\left(z+ z^{-1}\right)\right)
P_n^{-2}  \dex z
\label{eq:22}
\end{equation}
({\it the associated functions}). 
As opposed to $P_n$, 
it is obviously singular in the point 
$z=0$.
Thus, having expanded  $\tilde P_n$ through the basis  ${P_n,Q_n}$, 
it cannot involve any
fraction of $Q_n$ in its `content' and, thus,
$z^n(P'_n(z^{-1})-\mu P_n(z^{-1}))\propto P_n(z)$.
The constant proportionality 
coefficient can be fixed
evaluating the equations above at the
point
$z=1$.
In this way one may obtain
\begin{eqnarray}
  \label{eq:23}
P'_n(z)-\mu P_n(z)
&=&\epsilon (2\omega)^{-1}z^{n}
P_n(z^{-1}),
\end{eqnarray}
where $\epsilon^2=1$.
 
It is therefore shown that the fulfilment of 
\Eq (\ref{eq:23}) is the necessary condition for the 
polynomial $P_n(z)$ of degree $n$
to satisfy \Eq (\ref{eq:11}).
(In the general case, the formula (\ref{eq:21}) yields {\em the
  second\/}, linearly independent
solution of (\ref{eq:11})).
Conversely,
it is easy to prove that
if {\it any analytic function\/} obeys \Eq (\ref{eq:23})
then \Eq
(\ref{eq:11}) is also satisfied.

\myselector{\null}{
It is also worth noting the following representation of the phase
function through the polynomial $P_n$:
\begin{equation}
  \label{eq:3a}
  \exp(-\imi\ph(t))={\imi\epsilon}z^{n+1}{ P_n(z^{-1})\over  P_n(z)}.
\end{equation}
It follows from \Eqs
(\ref{eq:3}),(\ref{eq:6}),(\ref{eq:10}),(\ref{eq:23}).
}

\Eq (\ref{eq:23}) leads to the following 
relations mentioned above among the coefficients of $P_n$:
\begin{eqnarray}
  \label{eq:24a}
  \epsilon (2\omega)^{-1} a_0&=&-\mu a_n
\\
  \label{eq:24}
  \epsilon (2\omega)^{-1} a_k&=&
(n+1-k)a_{n+1-k}-\mu  a_{n-k},\;k=1,2\dots n,
\end{eqnarray}
where \Eq (\ref{eq:24a}) can be considered as 
a particular case of \Eq (\ref{eq:24}), provided 
one has introduced  $a_{n+1}\equiv0$.
As the above speculation claims, they are
equivalent to \Eq  (\ref{eq:11}). This implies some further
relationships discussed below.

The elements of the matrix $\mathbf{G}^{(\epsilon)}$
of the linear system (\ref{eq:24}),(\ref{eq:24a})  
can be represented in  terms of the Kronecker delta symbols as follows:
\begin{eqnarray}
  \label{eq:25}
G^{(\epsilon)}_{j\;k}&=&\epsilon(2\omega)^{-1}\delta_{j\;k} +
\mu \delta_{j \;n-k} -j\delta_{j\; n+1-k},\;j,k=0,1,\dots,n.
\end{eqnarray}
Then the product 
of the ({\it commuting}) matrices
$\mathbf{G}^{(+1)}$, $\mathbf{G}^{(-1)}$
is easily computable and one gets
\myselector{
$ 
\mathbf{G}^{(+1)}\cdot\mathbf{G}^{(-1)}=\mathbf{\Phi},
$ 
see (\ref{eq:14}).
}{
\begin{eqnarray}
  \label{eq:26}
(\mathbf{G}^{(+1)}\cdot\mathbf{G}^{(-1)})_{j\;l}&=&
(\lambda-j(n+1-j))\delta_{j\;l}
+(n-j)\mu \delta_{j\;l-1}
+j\mu\delta_{j\;l+1}.
\end{eqnarray}
The right-hand-side expression
here is nothing else but the component representation of
the matrix $\mathbf{\Phi}$  (\ref{eq:14}).
}
Thus $\mathbf{\Phi}$ is factorizable, provided 
the constraint 
$4\omega^2(\lambda+\mu^2)=1$ (see \Eqs (\ref{eq:11a}))
is taken into account.
The same concerns 
the determinants which are factorized as follows:
\myselector{
$  \Delta_n(\lambda,\mu)=
\det\mathbf{G}^{(+1)}\cdot
\det\mathbf{G}^{(-1)}. $
}
{
$$
 \Delta_n(\lambda,\mu)=
\det\mathbf{G}^{(+1)}\cdot
\det\mathbf{G}^{(-1)}.
$$
}
Therefore the spectral equation (\ref{eq:14})
is equivalent to the condition
\begin{equation}
  \label{eq:28}
  \mbox{either\ }\det\mathbf{G}^{(+1)}=0
\mbox{\ or\ }\det\mathbf{G}^{(-1)}=0,
\end{equation}
where the either branch involves the
algebraic equation restricting parameters $\mu,\omega$
(and depending on $n$).
This property is useful in numerical applications.

Finally, let us consider polynomial solutions of the {\em two\/}
specimens of \Eq   (\ref{eq:11}) with common parameter 
$\mu$ but different integer  $n$'s and $\lambda$'s obeying \Eq 
(\ref{eq:14a}) which we denote 
${n_1},{n_2},\lambda^{(1)},\lambda^{(2)}$.
An automatic computation establishes the
validity of the following identity
\begin{eqnarray}
  \label{eq:29}
0&=&
  {\dex\over\dex z}\left(
z^{-({n_1}+{n_2})/2}
\exp(-\mu(z+z^{-1}))\times
             \vphantom{{\dex P_{{n_1}}\over\dex z} }
\right.
\\&&\left.
\left[
P_{{n_2}}{\dex P_{{n_1}}\over\dex z}
-
P_{{n_1}}{\dex P_{{n_2}}\over\dex z}
-
\half({n_1}-{n_2})z^{-1}
P_{{n_1}} P_{{n_2}}
\right]\right)
+\Xi_{n_1,n_2} P_{{n_1}} P_{{n_2}},
\nonumber \\  \label{eq:30}
 \llap{\mbox{where\ }}\Xi_{n_1,n_2}&=&
 z^{-({n_1}+{n_2})/2}
 \exp(-\mu(z+z^{-1}))\times
\nonumber\\&&
 \left[
 \left(\lambda^{(1)}-\lambda^{(2)}
 -\quart
 ({n_1}-{n_2})({n_1}+{n_2}+2)\right)z^{-2}
\right.
\nonumber\\&&\left.\;\;
 +\half \mu ({n_1}-{n_2})  z^{-1}(1+z^{-2})
\right].
\end{eqnarray}
It implies the {\em theorem\/}:
\begin{quote}\em
For $\mu>0$
 the polynomial solutions 
of \Eq  (\ref{eq:11}) of different degrees $n_1,n_2$
are orthogonal on the semi-axis $\mathbb{R}^+$ with the weights
$\Xi_{n_1,n_2}$ (\ref{eq:30}), i.e.\
\begin{equation}
  \label{eq:31}
  \int_0^\infty \Xi_{n_1,n_2}  P_{{n_1}} P_{{n_2}} \dex z=0.
\end{equation}
\end{quote}
Under the same condition
the polynomials $P_n$ are also normalizable
in appropriate norm involving the factor $\exp(-\mu(z+z^{-1}))$.

\myselector{}{\bigskip}

Resuming, 
the nonlinear first order ODE (\ref{eq:1})
arising in the theory of Josephson junctions
with r.h.s.\ (\ref{eq:1a}) was shown to be equivalent 
to the linear homogeneous second order ODE 
with polynomial coefficients
(\ref{eq:8b}).
The latter equation was identified as a particular
instance of the double confluent
Heun equation. 
It means that 
{\em
the problem (\ref{eq:1}),(\ref{eq:1a}) 
proves completely solvable in terms of the double confluent Heun functions}.
The series of constraints on the problem parameters
enumerated by a non-negative integer parameter $n$ (see (\ref{eq:11a})) 
was
derived 
whose fulfilment leads to 
the existence of solutions 
representable in terms of polynomials ($P_n$)
 of degree $n$.
The corresponding master equation is (\ref{eq:11}),
the constraint equations 
are  
(\ref{eq:14a}), where  $\Delta_n$
can be computed by means of \Eq  (\ref{eq:19})
through the finite products of $2\times2$ matrices  with a single zero
element, 
the polynomial coefficients are determined by \Eq  (\ref{eq:20})
through the similar matrix products.
A curious first order linear ``non-classical'' two-argument
differential equation (\ref{eq:23}) which $P_n$ has to obey
was found. 
The polynomial solutions of \Eq   (\ref{eq:11})
were shown to
constitute the normalizable orthogonal system 
on the positive 
semi-axes $\mathbb{R}^+$.

\end{document}